\documentclass{ws-p8-50x6-00}
\usepackage{url}

\begin{document}

\title{WIMP searches with AMANDA-B10}

\address{\vspace*{-1ex}Presented by Joakim Edsj\"o for the AMANDA 
Collaboration:\\}

\author{
X.~Bai\lowercase{$^{a}$}, 
G.~Barouch\lowercase{$^{k}$}, 
S.W.~Barwick\lowercase{$^{h}$}, 
R.C.~Bay\lowercase{$^{g}$}, 
K.-H.~Becker\lowercase{$^{b}$}, 
L.~Bergstr\"om\lowercase{$^{m}$}, 
D.~Bertrand\lowercase{$^{c}$}, 
A.~Biron\lowercase{$^{d}$},
O.~Botner\lowercase{$^{l}$}, 
A.~Bouchta\lowercase{$^{d}$}, 
M.M.~Boyce\lowercase{$^{k}$}, 
S.~Carius\lowercase{$^{e}$}, 
A.~Chen\lowercase{$^{k}$},     
D.~Chirkin\lowercase{$^{7,2}$}, 
J.~Conrad\lowercase{$^{l}$}, 
J.~Cooley\lowercase{$^{k}$}, 
C.G.S.~Costa\lowercase{$^{c}$},
D.F.~Cowen\lowercase{$^{j}$}, 
J.~Dailing\lowercase{$^{h}$},
E.~Dalberg\lowercase{$^{m}$},  
T.~DeYoung\lowercase{$^{k}$}, 
P.~Desiati\lowercase{$^{d}$}, 
J.-P.~Dewulf\lowercase{$^{c}$}, 
P.~Doksus\lowercase{$^{k}$}, 
J.~Edsj\"o\lowercase{$^{m}$}, 
P.~Ekstr\"om\lowercase{$^{m}$},
T.~Feser\lowercase{$^{i}$}, 
M.~Gaug\lowercase{$^{d}$}, 
A.~Goldschmidt\lowercase{$^{f}$}, 
A.~Goobar\lowercase{$^{m}$},  
A.~Hallgren\lowercase{$^{l}$}, 
F.~Halzen\lowercase{$^{k}$},
K.~Hanson\lowercase{$^{j}$}, 
R.~Hardtke\lowercase{$^{k}$}, 
M.~Hellwig\lowercase{$^{i}$}, 
G.C.~Hill\lowercase{$^{k}$}, 
P.O.~Hulth\lowercase{$^{m}$}, 
S.~Hundertmark\lowercase{$^{h}$}, 
J.~Jacobsen\lowercase{$^{f}$}, 
A.~Karle\lowercase{$^{k}$}, 
J.~Kim\lowercase{$^{h}$}, 
B.~Koci\lowercase{$^{k}$},
L.~K\"opke\lowercase{$^{i}$}, 
M.~Kowalski\lowercase{$^{d}$}, 
J.I.~Lamoureux\lowercase{$^{f}$}, 
H.~Leich\lowercase{$^{d}$},  
M.~Leuthold\lowercase{$^{d}$}, 
P.~Lindahl\lowercase{$^{e}$}, 
P.~Loaiza\lowercase{$^{l}$}, 
D.M.~Lowder\lowercase{$^{g}$},
J.~Ludvig\lowercase{$^{f}$}, 
J.~Madsen\lowercase{$^{k}$},  
P.~Marciniewski\lowercase{$^{l}$},
H.S.~Matis\lowercase{$^{f}$},
T.C.~Miller\lowercase{$^{a}$},
Y.~Minaeva\lowercase{$^{m}$}, 
P.~Mio\v{c}inovi\'c\lowercase{$^{g}$}, 
R.~Morse\lowercase{$^{k}$}, 
T.~Neunh\"offer\lowercase{$^{i}$},
P.~Niessen\lowercase{$^{d}$}, 
D.R.~Nygren\lowercase{$^{f}$}, 
H.~Ogelman\lowercase{$^{k}$},
C.~P\'erez~de~los~Heros\lowercase{$^{l}$}, 
P.B.~Price\lowercase{$^{g}$}, 
K.~Rawlins\lowercase{$^{k}$},
W.~Rhode\lowercase{$^{b}$}, 
S.~Richter\lowercase{$^{d}$}, 
J.~Rodr\'\i guez~Martino\lowercase{$^{m}$}, 
P.~Romenesko\lowercase{$^{k}$}, 
D.~Ross\lowercase{$^{h}$}, 
H.~Rubinstein\lowercase{$^{m}$},
H.-G.~Sander\lowercase{$^{i}$}, 
T.~Scheider\lowercase{$^{i}$}, 
T.~Schmidt\lowercase{$^{d}$}, 
D.~Schneider\lowercase{$^{k}$}, 
R.~Schwarz\lowercase{$^{k}$}, 
A.~Silvestri\lowercase{$^{2,4}$}, 
G.~Smoot\lowercase{$^{f}$}, 
M.~Solarz\lowercase{$^{g}$}, 
G.M.~Spiczak\lowercase{$^{a}$}, 
C.~Spiering\lowercase{$^{d}$}, 
N.~Starinsky\lowercase{$^{k}$},
D.~Steele\lowercase{$^{k}$}, 
P.~Steffen\lowercase{$^{d}$}, 
R.G.~Stokstad\lowercase{$^{f}$}, 
O.~Streicher\lowercase{$^{d}$},
I.~Taboada\lowercase{$^{j}$}, 
M.~Vander~Donckt\lowercase{$^{c}$}, 
C.~Walck\lowercase{$^{m}$}, 
C.~Weinheimer\lowercase{$^{i}$}, 
C.H.~Wiebusch\lowercase{$^{d}$}, 
R.~Wischnewski\lowercase{$^{d}$}, 
K.~Woschnagg\lowercase{$^{g}$},
W.~Wu\lowercase{$^{h}$}, 
G.~Yodh\lowercase{$^{h}$} and 
S.~Young\lowercase{$^{h}$}
}

\address{
\begin{itemize} \setlength{\itemsep}{0cm}\setlength{\parskip}{0cm}
    \setlength{\parsep}{0cm}
   \item[(a)] Bartol Research Institute, Univ. of Delaware, Newark, DE 19716, USA
   \item[(b)] Fachbereich 8 Physik, BUGH Wuppertal, D-42097 Wuppertal, Germany
   \item[(c)] Brussels Free Univ., Science Faculty CP230, Boulevard du Triomphe, B-1050 Brussels, Belgium
   \item[(d)] DESY-Zeuthen, D-15735 Zeuthen, Germany
   \item[(e)] Dept. of Technology, Kalmar Univ., SE-39129 Kalmar, Sweden
   \item[(f)] Lawrence Berkeley National Laboratory, Berkeley, CA 94720, USA
   \item[(g)] Dept. of Physics, Univ. of California, Berkeley, CA 94720, USA
   \item[(h)] Dept. of Physics and Astronomy, Univ. of California, Irvine, CA 92697, USA
   \item[(i)] Inst. of Physics, Univ. of Mainz, Staudinger Weg 7, D-55099 Mainz, Germany
   \item[(j)] Dept. of Physics and Astronomy, Univ. of Pennsylvania, Philadelphia, PA 19104, USA
   \item[(k)] Dept. of Physics, Univ. of Wisconsin, Madison, WI 53706, USA
   \item[(l)] Dept. of Radiation Sciences, Uppsala Univ., SE-75121 Uppsala, Sweden
   \item[(m)] Dept. of Physics, Stockholm Univ., SE-11385 Stockholm, Sweden
\end{itemize}
}


\maketitle

\abstracts{ We report on the search for nearly vertical up-going muon
neutrinos from WIMP annihilations in the center of the Earth with the
AMANDA-B10 detector.  The whole data sample collected in 1997, 10$^9$
events, has been analyzed and a final sample of 15 up-going events is
found in a restricted zenith angular region where a signal from WIMP
annihilations is expected.  A preliminary upper limit at 90\%
confidence level on the annihilation rate of WIMPs in the center of
the Earth is presented.}

\section{Introduction}

There are strong indications for the existence of dark matter in the
Universe.\cite{Bergstrom:00a}  The favourite
candidate for the dark matter is a Weakly Interacting Massive
Particle, a WIMP, of which the neutralino that arises in
supersymmetric extensions of the standard model is a natural
candidate.\cite{Jungman:96a}  If these WIMPs exist they accumulate in the center of the
Earth and the Sun, where they can annihilate pair-wise producing muon
neutrinos that can be searched for with neutrino telescopes like
AMANDA. Searches of this kind have been performed by existing
experiments like MACRO~\cite{MACRO:99a}, Baikal~\cite{Baikal:99a},
Baksan~\cite{Baksan:99a}, Kamiokande and
Super-Kamiokande~\cite{SuperK:00a}.  We here report on searches
for muon neutrinos from WIMP annihilations in the center of the Earth
with the AMANDA-B10 detector using data from 1997.

\section{The AMANDA-B10 detector}

AMANDA-B10 consists of an array of 302 Optical Modules (OM) arranged
in 10 strings deployed between 1500 and 2000 meters in the South Pole
ice cap.  Muons from charged-current neutrino interactions near the
array are detected by the Cherenkov light they emit when traversing
the ice.  The timing of the Cherenkov photons reaching the OMs enables
us to reconstruct the muon track.  A more detailed description of the
detector is given in~\cite{Amanda:00b}.

\section{Signal and background simulations}

Our main backgrounds when searching for a WIMP signal are the
atmospheric neutrinos and muons.  We first discuss the signal and then the
background simulations.

\subsection{Simulation of WIMP annihilations}

WIMPs annihilate pair-wise to, e.g., leptons, quarks and gauge and 
Higgs bosons.  High-energy neutrinos are produced in the decays
and/or hadronization of these annihilation products.  Neutrinos
produced in quark jets (from e.g.\ $b\bar{b}$ or Higgs bosons)
typically have lower energy than those produced from decays of $\tau$
leptons and gauge bosons.  We will refer to the first type of
annihilation channels as `soft' and to the second as `hard'.  As a
typical soft channel we choose $b \bar{b}$ and as a typical hard
channel we choose $W^{+}W^{-}$ above the $W^{+}W^{-}$ threshold and
$\tau^{+}\tau^{-}$ below.  The hadronization and/or decay of the
annihilation products were simulated with {PYTHIA}.\cite{PYTHIA}
For details, see \cite{Bergstrom:98a}.

\subsection{Simulation of the atmospheric neutrino flux}

We have simulated the expected atmospheric neutrino flux using the
calculations of Lipari~\cite{Lipari:93a} and the neutrino and
anti-neutrino$-$nucleon cross sections from Gandhi et
al.\cite{Gandhi:96a}  The actual neutrino-nucleon interactions have
been simulated with {PYTHIA}~\cite{PYTHIA} using the
{CTEQ3}~\cite{CTEQ3} parameterization of the nucleon structure
functions.
A 3-year equivalent atmospheric neutrino sample with energies between
10~GeV and 10~TeV and zenith angles between 90$^\circ$ and 180$^\circ$
has been simulated.\cite{Dalberg:99a}

\subsection{Simulation of the atmospheric muon flux}

The majority of the triggers in AMANDA are induced by muons produced
in cosmic ray interactions in the atmosphere and reaching the detector
depth.  The simulation of this atmospheric muon flux was performed
with an algorithm described in ~\cite{BASIEV}.  Incoming protons were
generated between 0$^\circ<\theta<$85$^\circ$ and with energies
between 1.3~TeV and 1000~TeV assuming a differential energy
distribution of $E^{-2.67}$.   Simulating a statistically significant sample of atmospheric
muon background is an extremely CPU-intensive task due to the high
rejection factors needed.  We have simulated $6.3\times 10^{10}$
primary interactions, giving $3.9 \times 10^{7}$ atmospheric muon
triggers, which corresponds to 8 days of detector lifetime.

In all these simulations, muon interaction probabilities
from~\cite{Lohmann:85a} were used.

\begin{figure}
\centering\epsfig{file=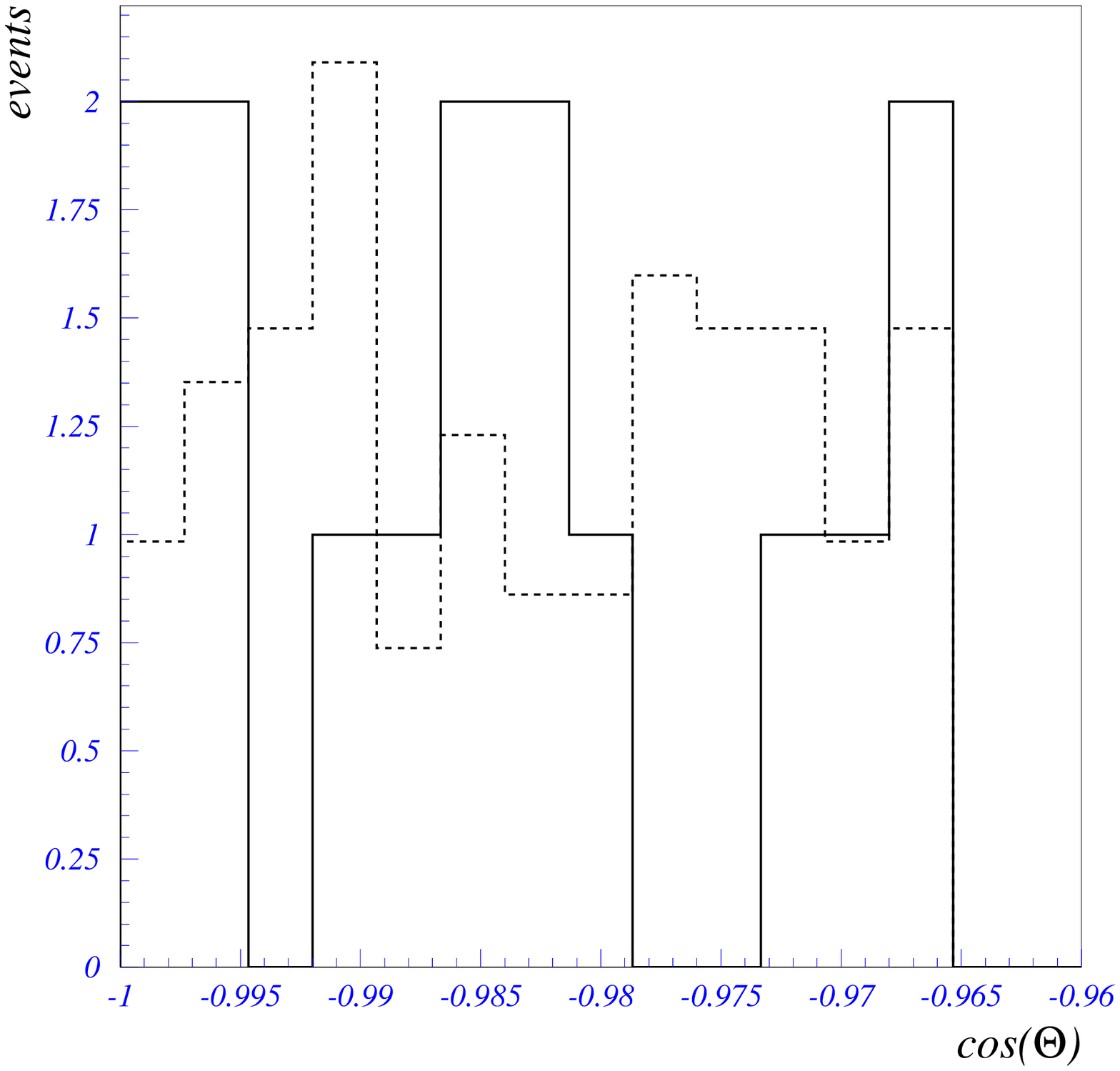,width=0.49\textwidth}
\centering\epsfig{file=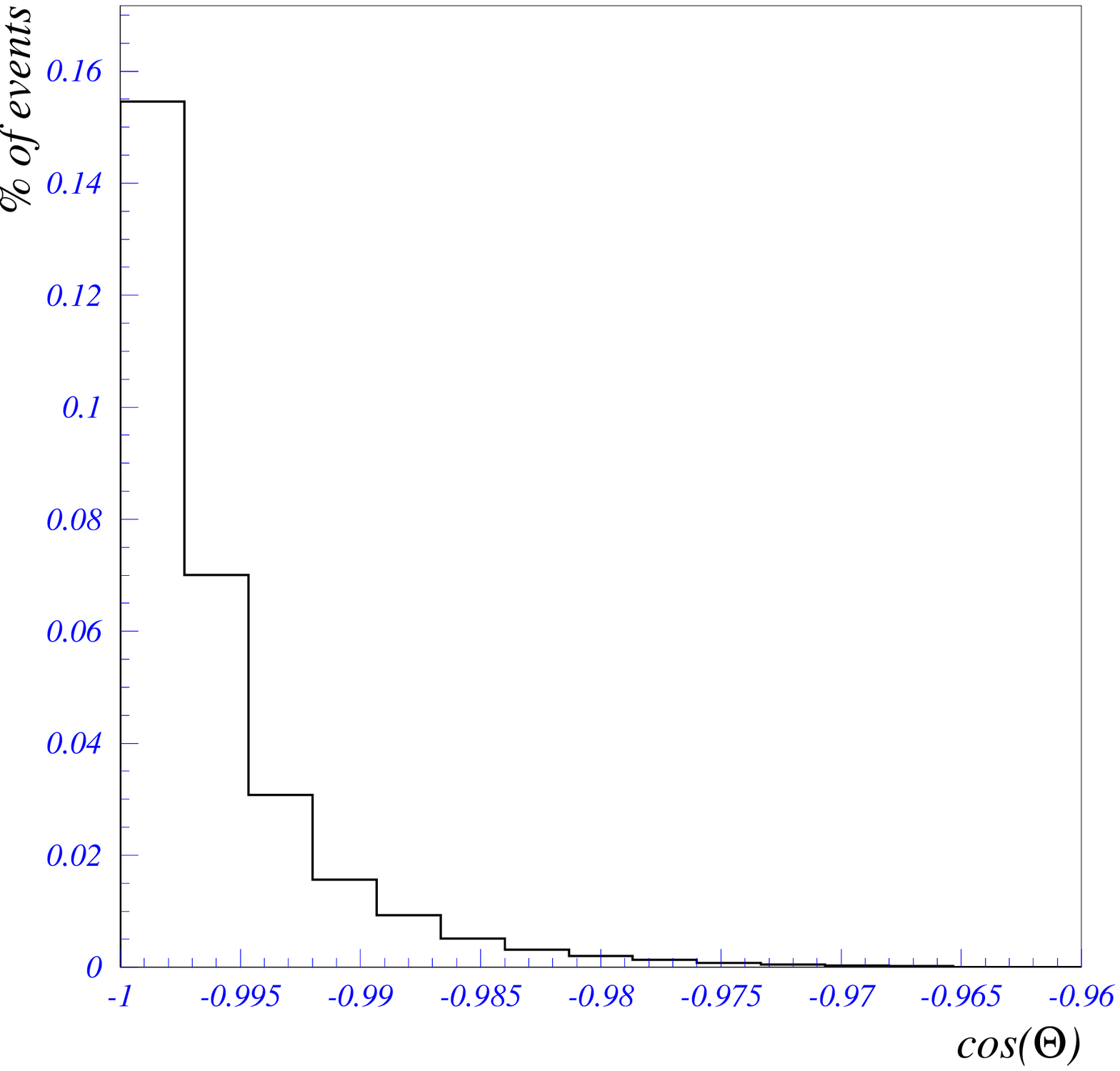,width=0.49\textwidth}
\caption{ {\bf Left:} Angular distribution of the remaining data (full line) 
and simulated atmospheric neutrino  events (dashed line) after filter level
5. The angular range is between 165$^\circ$ and 180$^\circ$. 
The atmospheric neutrino sample has 
been normalized to the 135 days of lifetime of the detector in 1997.
{\bf Right:} Angular distribution of the remaining fraction of
simulated WIMP signal (mass 250 GeV with hard annihilation channel) 
at filter level 5.}
\label{fig:L5_costh}
\end{figure}

\section{Data analysis}

The analysis presented in this paper was performed on data taken with
the 10-string AMANDA detector between March and November 1997.  The
original data set consists of 10$^9$ events in a total of 135 days of
detector lifetime.  We apply five different
levels of cuts which primarily cut on the zenith angle, the number of
hits, the track length and some variables that describe the space and
time topology of the event.\cite{Pia:00a,ama-wimp}
At level 5, we have cut away all the atmospheric muon background and 
are left with 15 neutrino events in data which is consistent with the 
expected 16.6 events from the atmospheric neutrinos. The passing rate 
for a WIMP with mass 250 GeV and a hard annihilation spectrum is 29\% 
at the same level.

\section{Results}
\label{sec:results}

In Fig.~\ref{fig:L5_costh} we plot the events at level 5 together with
the prediction for the atmospheric neutrinos.\cite{Pia:00a}  We also
plot the expected angular distribution from a WIMP. As can be clearly
seen, we do not see any statistically significant excess of nearly
vertical neutrino-induced muons.  Hence, we will use these remaining
events to derive an upper limit on the WIMP signal.

\subsection{Limits on the WIMP signal}

\begin{table}
{\small
\begin{center}
\begin{tabular}{lrcccc}     \hline
$M_{\chi}$ &Ann. & Optimal  & $\Gamma_{\nu\mu}$ & 
 $\Gamma_{A}$ & $\Phi_{\mu}$ \\
 {}[GeV]          &ch.  & $\theta_{\rm cut}$ & [10$^{-20}$ cm$^{-3}$ s$^{-1}$] & 
 [s$^{-1}$]   & [km$^{-2}$ yr$^{-1}$] \\ \hline

100& hard         & $170^{\circ}$   & 134   &  2.6$\times$10$^{14}$
  & 5\,700 \\
   &soft          & $165^{\circ}$   & 1338  &  1.3$\times$10$^{16}$ 
  & 42\,000 \\ \hline
250& hard         & $172.5^{\circ}$ & 14.1  &  9.7$\times$10$^{12}$ 
  & 1\,500 \\
   &soft          & $172.5^{\circ}$ & 89    &  2.2$\times$10$^{14}$ 
  & 4\,000\\ \hline
500& hard         & $172.5^{\circ}$ & 5.7   &  1.9$\times$10$^{12}$ 
  & 1\,200 \\
   & soft         & $172.5^{\circ}$ & 27    &  3.1$\times$10$^{13}$ 
  & 2\,000 \\ \hline  
1000& hard        & $172.5^{\circ}$ & 2.8   &  4.6$\times$10$^{11}$
  & 1\,000 \\
    & soft        & $172.5^{\circ}$ & 11.5  &  7.0$\times$10$^{12}$ 
  & 1\,400 \\ \hline 
3000&  hard       & $172.5^{\circ}$ & 1.2   &  6.4$\times$10$^{10}$ 
  & 960 \\
    &soft         & $172.5^{\circ}$ & 4.3   &  1.1$\times$10$^{12}$
  & 1\,100 \\ \hline
5000& hard        & $172.5^{\circ}$ & 0.9   &  2.7$\times$10$^{10}$ 
  & 930 \\
     &soft        & $172.5^{\circ}$ & 3.1   &  5.5$\times$10$^{11}$ 
  & 1\,100 \\\hline
\end{tabular}
\end{center}
}
\caption{ The 90$\%$ confidence level on the neutrino to muon
conversion rate ($\Gamma_{\nu\mu}$) (for muons above 10 GeV), 
the WIMP annihilation rate ($\Gamma_{A}$) and the muon flux (above 1 
GeV)\protect\cite{Pia:00a}.}
\label{tab:Gamma_numu}
\end{table}

As seen in Fig.~\ref{fig:L5_costh}, the WIMP signal is very peaked
towards vertical muons, suggesting that we should cut further in the
zenith angle.  If we cut at $\theta=165^{\circ}, 170^{\circ}, 
172.5^{\circ}$ or $175^{\circ}$ we can derive 90\% C.L. 
limits on the number of signal events, 7.0, 5.6, 4.5 and 5.2 events 
respectively. Although the systematic uncertainties for different 
angular cuts currently are under investigation, we choose here to use 
the angular window which gives the best limits.
{}From these limits on the number of signal events we can
calculate limits on the neutrino-to-muon conversion rate, $\Gamma_{\mu
\nu}$, which is the number of neutrinos that interact and produce a
muon per volume element per second.  In Table~\ref{tab:Gamma_numu} we
show the limits on the neutrino-to-muon conversion rate where the muon
energy (at the neutrino-nucleon vertex) is above 10 GeV\@.  We also
show the corresponding limits on the WIMP annihilation rate in the
center of the Earth, $\Gamma_{A}$ and to compare with previous
experiments, we also show the calculated limits on the
neutrino-induced muon flux above 1 GeV\@.

\begin{figure}
\begin{minipage}[t]{0.48\textwidth}
\centering\epsfig{file=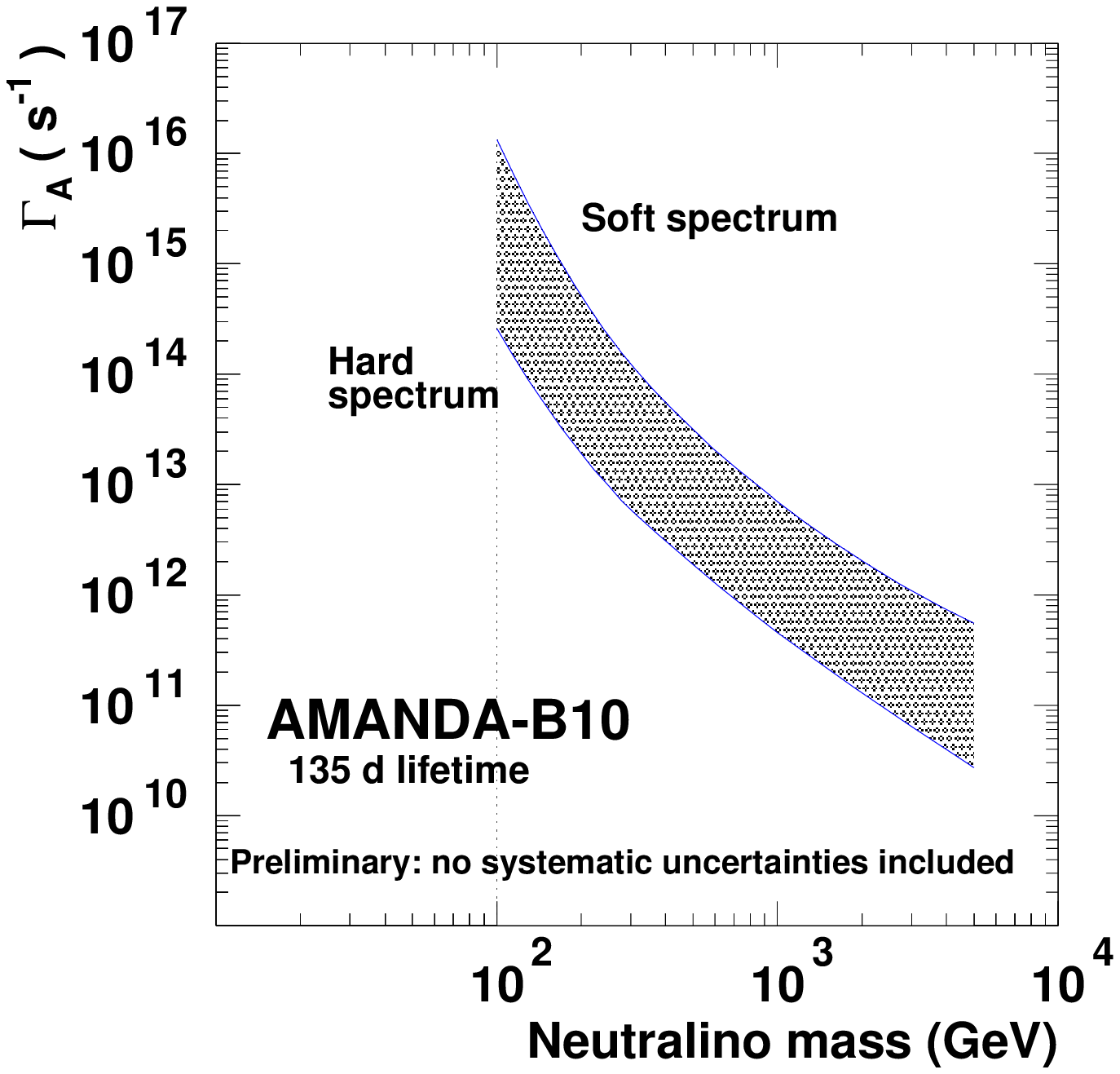,width=\textwidth}
\caption{90\% confidence level upper limits on the WIMP
annihilation rate in the center of the Earth, as a function of the 
WIMP mass and for the two extreme annihilation channels
considered in the analysis.}
\label{fig:annrate_limits}
\end{minipage}\hfill
\begin{minipage}[t]{0.48\textwidth}
\centering\epsfig{file=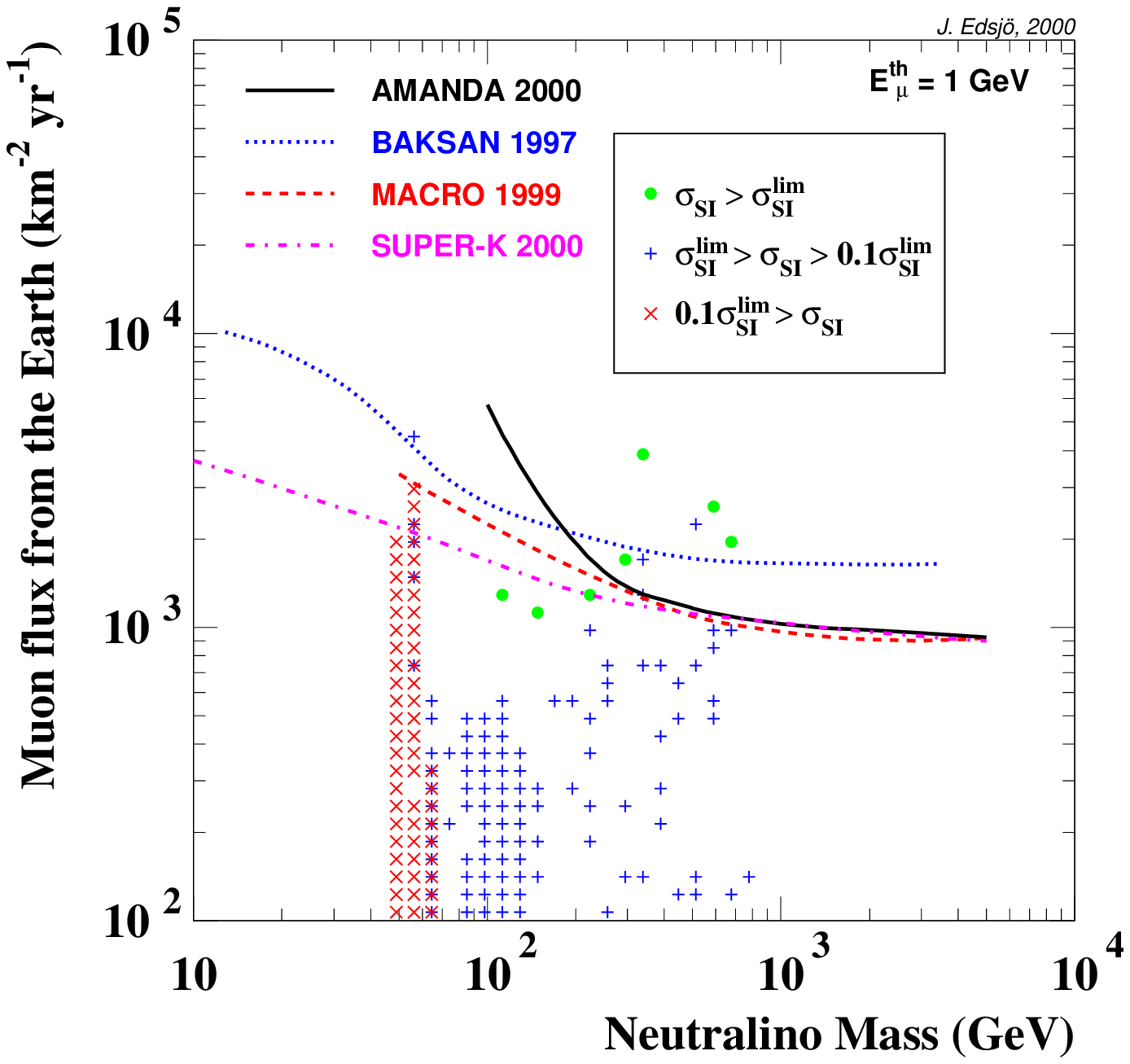,width=\textwidth}
\caption{90\% confidence level upper limits on the 
neutrino-induced muon flux for a hard annihilation spectrum. 
The limit on the muon flux has been 
calculated from the limits on the annihilation rate with a muon 
energy threshold of 1 GeV to be able to compare with other 
experiments. Also shown are predictions from calculations in the 
MSSM.\protect\cite{Bergstrom:98a}}
\label{fig:muflux_limits}
\end{minipage}
\end{figure}

In Fig.~\ref{fig:annrate_limits}, the limits on the annihilation rate
are shown and in Fig.~\ref{fig:muflux_limits} we show the limits on the
neutrino-induced muon flux.  In that figure we also show predictions
from the MSSM \cite{Bergstrom:98a} when the WIMP is a neutralino and 
we also compare with other experiments.  We
see that, especially at higher masses, AMANDA is already with this
limited data set competitive with previous experiments.

\subsection{Systematic uncertainties}

The limits are subject to experimental and theoretical systematic
uncertainties which affect the estimated background and the effective
volume and propagate to our derived limits.  The precise effect is
currently under detailed investigation but preliminary estimates
indicate that the limits may have to be increased by about a factor of
2--3.\cite{ama-wimp}

\section{Conclusions}

We have performed a search for an excess of vertically upgoing muons
with the AMANDA-B10 neutrino detector as a signature for WIMP
annihilations in the center of the Earth.  Limits on the signal from
WIMPs have been derived from the non-observation of an effect. 
The limits presented here are obtained with 135 days of effective
detector lifetime and are already comparable with limits obtained by
Baikal, Baksan, Super-Kamiokande and MACRO, with much longer
accumulated lifetimes.  The systematic uncertainties are under
investigation and preliminary results suggest that the limits may
increase by about a factor of 2--3 when they are included.

\section*{Acknowledgments}

{\footnotesize
This research was supported by the U.S. NSF office of Polar Programs
and Physics Division, the U. of Wisconsin Alumni Research Foundation,
the U.S. DoE, the Swedish Natural Science Research Council, the
Swedish Polar Research Secretariat, the Knut and Alice Wallenberg
Foundation, Sweden, the German Ministry for Education and Research,
the US National Energy Research Scientific Computing Center (supported
by the U.S. DoE), U.C.-Irvine AENEAS Supercomputer Facility, and
Deutsche Forschungsgemeinschaft (DFG).  D.F.C. acknowledges the
support of the NSF CAREER program.  P. Desiati was supported by the
Koerber Foundation (Germany).  C.P.H. received support from the EU 4th
framework of Training and Mobility of Researchers.  St. H. is
supported by the DFG (Germany).  P. Loaiza was supported by a grant
from the Swedish STINT program.
}

\end{document}